\documentclass{Interspeech2024}
\usepackage{multirow}
\usepackage{graphicx}
\usepackage{subcaption}
\interspeechcameraready

\title{Noise-robust Speech Separation with Fast Generative Correction}

\name[affiliation={1}]{Helin}{Wang}
\name[affiliation={1,2}]{Jesús}{Villalba}
\name[affiliation={1}]{Laureano}{Moro-Velazquez}
\name[affiliation={3}]{Jiarui}{Hai}
\name[affiliation={1}]{Thomas}{Thebaud}
\name[affiliation={1,2}]{Najim}{Dehak}
\address{
  $^1$Center for Language and Speech Processing, Johns Hopkins University, USA\\
  $^2$Human Language Technology Center of Excellence, Johns Hopkins University, USA \\
  $^3$Laboratory for Computational Auditory Perception, Johns Hopkins University, USA}
\email{hwang258@jhu.edu}




\begin{document}

\maketitle

\begin{abstract}
Speech separation, the task of isolating multiple speech sources from a mixed audio signal, remains challenging in noisy environments. 
In this paper, we propose a generative correction method to enhance the output of a discriminative separator. 
By leveraging a generative corrector based on a diffusion model, we refine the separation process for single-channel mixture speech by removing noises and perceptually
unnatural distortions. 
Furthermore, we optimize the generative model using a predictive loss to streamline the diffusion model's reverse process into a single step and rectify any associated errors by the reverse process. 
Our method achieves state-of-the-art performance on the in-domain Libri2Mix noisy dataset, and out-of-domain WSJ with a variety of noises, improving SI-SNR by 22-35\% relative to SepFormer, demonstrating robustness and strong generalization capabilities. % to out-of-domain data.
\end{abstract}

\section{Introduction}

Speech separation, also known as the cocktail party problem, is a foundational problem in speech and audio processing, garnering significant attention in research~\cite{wang2014training}. This paper focuses on single-channel two-speaker speech separation in noisy environments, which is much more challenging than clean speech separation~\cite{Wichern2019WHAM}.

In recent years, deep learning models have emerged as powerful tools for addressing speech separation, exhibiting notable performance gains over traditional methods~\cite{wang2018supervised}.
Hershey \textit{et al.} introduced a clustering method leveraging trained speech embeddings for separation~\cite{hershey2016deep}. Yu \textit{et al.} proposed Permutation Invariant Training (PIT) at the frame level for source separation~\cite{yu2017permutation}. Luo \textit{et al.} pioneered an influential deep learning method for speech separation in the time domain, employing an encoder, separator, and decoder~\cite{luo2019conv}. Following this structure, various discriminative models have been developed, such as WaveSplit~\cite{zeghidour2021wavesplit}, DPTNet~\cite{chen20l_interspeech}, SepFormer~\cite{subakan2021attention}, TF-GridNet~\cite{wang2023tf} and MossFormer~\cite{zhao2023mossformer}.
These models directly optimize the Scale-Invariant Signal-to-Noise Ratio (SI-SNR) metric~\cite{hershey2016deep}, 
and the performance on WSJ0-2mix~\cite{hershey2016deep} has been largely saturated~\cite{lutati2022sepit}.
However,
they cannot generalize well to speech coming from a wider range of speakers and recorded in slightly different conditions, such as noisy and reverberant environments~\cite{maciejewski2020whamr,cosentino2020librimix}.
Furthermore, the SI-SNR metric does not perfectly align with human perception, as models trained using it may introduce perceptually unnatural distortions that could adversely affect downstream tasks~\cite{le2019sdr}.

In contrast to discriminative models, generative models aim to learn a prior distribution of the data. Recently, score-based generative models~\cite{song2021scorebased}, also referred to as diffusion models, have been introduced for various speech processing tasks, such as speech enhancement and dereverberation~\cite{lu2022conditional,richter2023speech,lemercier2023storm}, target speech extraction~\cite{kamo23_interspeech,hai2023dpm} and speech separation~\cite{scheibler2023diffusion,hirano2023diffusionbased,lutati2024separate,liu2024generative}.
While these methods are capable of generating audio samples with good perceived quality and have the potential to generalize well to out-of-domain data, 
they struggle with the permutation problem of speakers~\cite{yu2017permutation} and often yield inferior results on reference-based measurements~\cite{hirano2023diffusionbased}. 
Additionally, such diffusion models require multiple reverse steps, which can significantly increase computational time for inference.

To leverage the strengths of both discriminative and generative models, 
we introduce a method called \textbf{Ge}nerative \textbf{Co}rrection (GeCo\footnote{Source code and pretrained models are available at: \href{https://github.com/WangHelin1997/Fast-GeCo}{https://github.com/WangHelin1997/Fast-GeCo}. Audio Samples are available at: 
\href{https://fastgeco.github.io/Fast-GeCo/}{https://fastgeco.github.io/Fast-GeCo/}.
}) for noise-robust speech separation. 
GeCo utilizes a corrector based on a diffusion model to refine the output of a discriminative separator. 
The separator predicts the initial separation and solves the permutation of speakers during training, 
while the corrector removes noises and perceptually unnatural distortions introduced by the separator.
Furthermore, to reduce the reverse steps of the diffusion model during inference and mitigate the discretization error resulting from the reverse process, 
we propose to fine-tune the GeCo model by optimizing a predictive loss, transforming the reverse process into a single step (referred to as Fast-GeCo).
Previous works by Lutati et al.~\cite{lutati2024separate} and Hirano et al.~\cite{hirano2023diffusionbased} also proposed refiners based on diffusion models for speech separation. 
However, these methods suffer from slow inference speed and did not demonstrate advancements in noisy speech separation.
We evaluate our proposed method on the Libri2Mix noisy dataset~\cite{cosentino2020librimix}, 
and the results demonstrate a significant improvement in separation performance, 
surpassing other state-of-the-art methods. 
In addition, experiments on different out-of-domain noisy data show notable enhancements in perceptual quality with GeCo and Fast-GeCo.

\begin{figure*}[t]
    \centering
    \begin{subfigure}{0.32\textwidth}
        \centering
        \includegraphics[width=\linewidth]{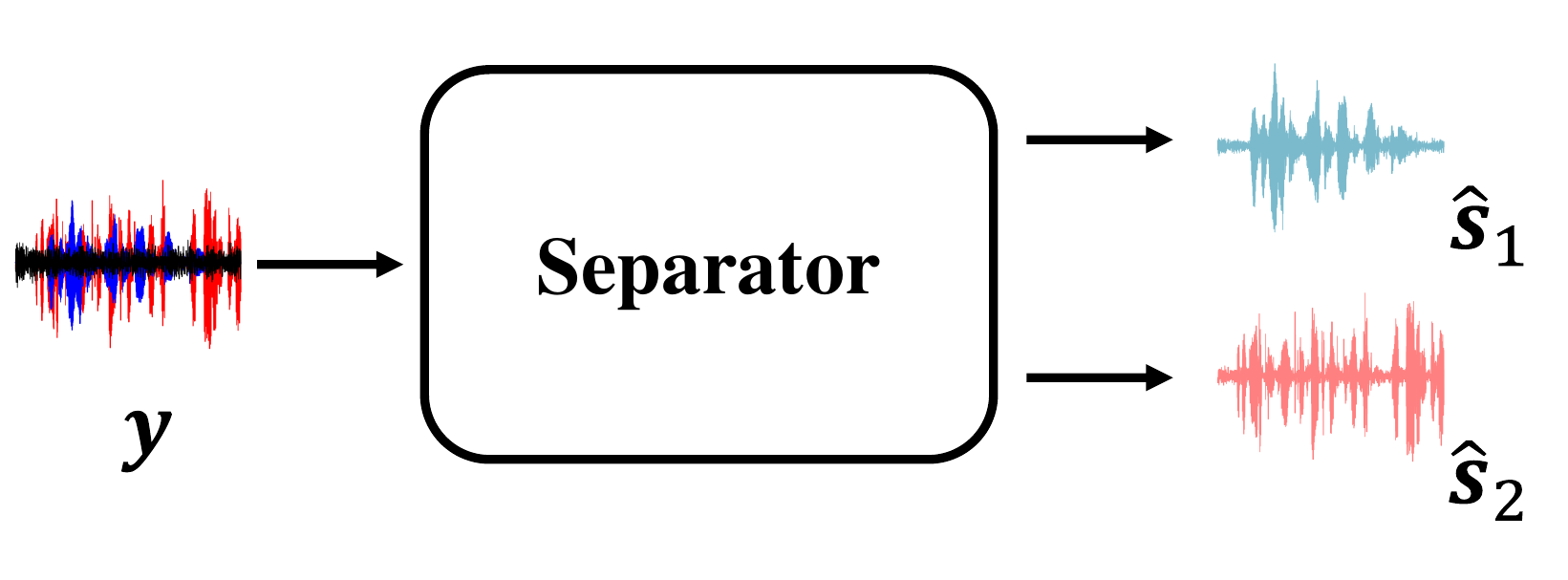}
        \caption{Discriminative separator.}
        \label{fig:sub1}
    \end{subfigure}
    \begin{subfigure}{0.32\textwidth}
        \centering
        \includegraphics[width=\linewidth]{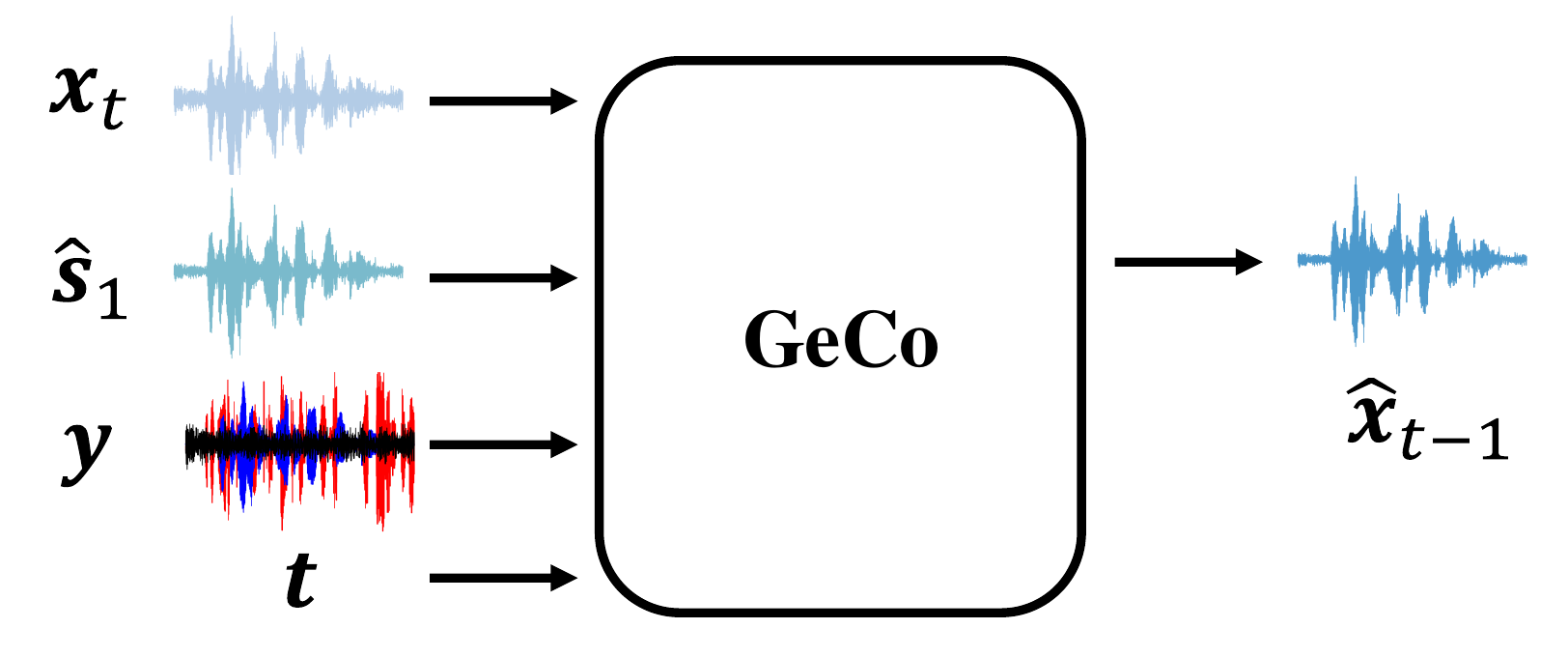}
        \caption{Generative corrector.}
        \label{fig:sub2}
    \end{subfigure}
    \begin{subfigure}{0.32\textwidth}
        \centering
        \includegraphics[width=\linewidth]{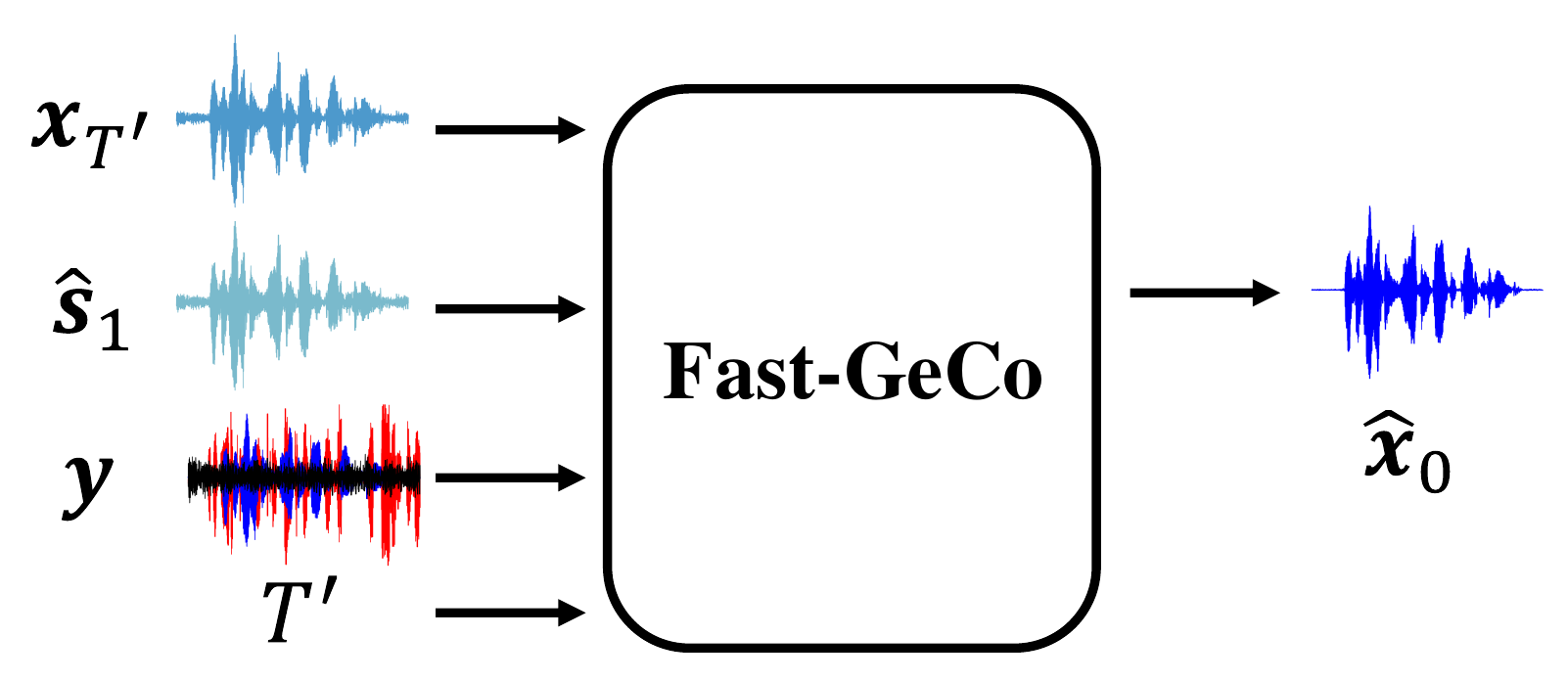}
        \caption{Fast generative corrector.}
        \label{fig:sub3}
    \end{subfigure}
    \vspace{-2mm}
    \caption{Illustration of the discriminative separator, generative corrector and fast generative corrector. Here, we use two-speaker separation as an example, and (b) and (c) depict an example of the first speaker. Three models are trained in order. During inference, we can use the discriminative separator followed by either the generative corrector or the fast generative corrector.}
    \label{fig:all}
    \vspace{-3mm}
\end{figure*}

\section{Background}

\subsection{Notation and Signal Model}
% Let bold lower and upper case letters represent vectors and matrices, respectively. 
The task of noisy speech separation is to extract different
speakers’ utterances and route them to different output signals $\boldsymbol{s}_k \in \mathbb{R}^N$ from a noisy mixture $\boldsymbol{y} \in \mathbb{R}^N$,
\begin{equation}
    \label{eq0}
    \boldsymbol{y} = \boldsymbol{n} + \sum_{k=1}^{K} \boldsymbol{s}_k
\end{equation}
where $\boldsymbol{n} \in \mathbb{R}^N$ is backgound noise and $N$ is the number of data points. $K$ is the number of sources, which is 2 in this paper.

\subsection{Separator}
A discriminative separator $f_{\text{sep}}$ is trained to directly minimize the difference between the estimated signals and the ground-truth signals.
Following~\cite{luo2019conv}, the SI-SNR loss is used to optimize the parameters of the separator.
\begin{equation}
\label{eq1}
\mathcal{L}_{\text{sep}}:= -10 \sum_{k=1}^{K} \log _{10} \frac{\Vert \frac{\langle\hat{\mathbf{s}}_k, \mathbf{s}_k\rangle \mathbf{s}_k}{\Vert\mathbf{s}_k\Vert^2} \Vert^2}{\Vert\hat{\mathbf{s}}_k-\frac{\langle\hat{\mathbf{s}}_k, \mathbf{s}_k\rangle \mathbf{s}_k}{\Vert\mathbf{s}_k\Vert^2}\Vert^2}
\end{equation}
where $\hat{\mathbf{s}}_k = f_{\text{sep}}(\boldsymbol{y};\theta_{\text{sep}}) \in \mathbb{R}^N$ is the estimated $k$-th clean signal, 
$\theta_{\text{sep}}$ are trainable parameters of the separator,
and $\Vert\mathbf{s}\Vert^2=\langle\mathbf{s}, \mathbf{s}\rangle$ denotes the signal power.
Scale invariance is maintained by normalizing $\hat{\mathbf{s}}_k$ and $\boldsymbol{s}_k$ to zero mean before computation. 
During training, utterance-level Permutation Invariant Training (uPIT) is employed to resolve the source permutation issue~\cite{kolbaek2017multitalker}.

\subsection{Stochastic Differential Equations (SDE)}
Different from discriminative models, generative models approximate complex data distributions, often showing better generalization ability to out-of-domain data and producing more natural sounding speech~\cite{lutati2024separate, lu2022conditional}. Initial experiments reveal that traditional discriminative separators do not generalize well to out-of-domain noisy data and may introduce perceptually unnatural artifacts. To achieve better separation, a diffusion probabilistic model is applied to modify the output of the separator.

Following Song \textit{et al.}~\cite{song2021scorebased}, 
we devise a stochastic diffusion process $\left\{\mathbf{x}_t\right\}_{t=0}^T$ indexed by a continuous time variable $t \in[0, T]$, which is represented as the solution to a linear SDE.
\begin{equation}
\label{eq2}
\mathrm{d} \mathbf{x}_t=\mathbf{f}\left(\mathbf{x}_t, \hat{\mathbf{s}}_1\right) \mathrm{d} t+g(t) \mathrm{d} \overrightarrow{\mathbf{w}}
\end{equation}
Here, $\overrightarrow{\mathbf{w}}$ is the standard Wiener process (\textit{a.k.a.}, Brownian motion), $\mathbf{f}\left(\mathbf{x}_t, \hat{\mathbf{s}}_1\right): \mathbb{R}^N\rightarrow \mathbb{R}^N$ is the drift coefficient of $\mathbf{x}_t$, and $g(\cdot): \mathbb{R}\rightarrow \mathbb{R}$ is the diffusion coefficient of $\mathbf{x}_t$.
$\mathbf{x}_0$ denotes the initial state of the speech, \textit{i.e.}, $\mathbf{s}_1$ (we use the first clean speech signal as an example), 
and $\mathbf{x}_T$ denotes the final state, \textit{i.e.}, the estimated speech signal by the separator $\hat{\mathbf{s}}_1$. Thus, this SDE diffuses the clean sample $\mathbf{s}_1$ into the noisy sample $\hat{\mathbf{s}}_1$, estimated by the separator.

We take the definition of Brownian Bridge SDE~\cite{lay2023reducing,qiu2023se}.
The BBED drift and diffusion coefficients are given by
\begin{equation}
\label{eq3}
\mathbf{f}\left(\mathbf{x}_t, \hat{\mathbf{s}}_1\right)=\frac{\hat{\mathbf{s}}_1-\mathbf{x}_t}{1-t}
\end{equation}
\begin{equation}
\label{eq4}
g(t)=c v^t
\end{equation}
where $c, v>0$.
To avoid division by zero in~(\ref{eq3}), $T$ is set slightly smaller than 1. By solving~\eqref{eq2} with~\eqref{eq3} and~\eqref{eq4}, we find that
the process state $\mathbf{x}_t$ follows a Gaussian perturbation kernel,
\begin{align}
\label{eq5}
\mathbf{x}_t&=\boldsymbol{\mu}\left(\mathbf{x}_0, \hat{\mathbf{s}}_1, t\right)+\sigma(t) \mathbf{z}_t\\
\label{eq6}
\boldsymbol{\mu}\left(\mathbf{x}_0, \hat{\mathbf{s}}_1, t\right) &=(1-t) \mathbf{x}_0+t \hat{\mathbf{s}}_1
\end{align}
where $\mathbf{z}_{t} \sim \mathcal{N}(\mathbf{0}, \mathbf{I})$ is random Gaussian noise.
Following~\cite{lay2023reducing},
\begin{align}
    \label{eq7}
    \sigma(t) &=\sqrt{(1-t) c^2\left[\left(v^{2 t}-1+t\right)+\log \left(v^{2 v^2}\right)(1-t) E\right]} \\
    \label{eq8}
    E&=\mathrm{Ei}[2(t-1) \log (v)]-\operatorname{Ei}[-2 \log (v)]
\end{align}
where $\text{Ei}(x) = \int_{-\infty}^{x} \frac{e^t}{t} dt$ is the exponential integral function~\cite{bender2013advanced}.

Following Song \textit{et al.}~\cite{song2021scorebased}, 
the SDE in~(\ref{eq2}) has an associated reverse SDE, which allows us to re-generate the clean signal $\mathbf{s}_1$ given the separator estimate $\hat{\mathbf{s}}_1$,
\begin{equation}
\label{eq9}
\mathrm{d} \mathbf{x}_t=\left[-\mathbf{f}\left(\mathbf{x}_t, \hat{\mathbf{s}}_1\right)+g(t)^2 \nabla_{\mathbf{x}_t} \log p_t\left(\mathbf{x}_t \mid \hat{\mathbf{s}}_1\right)\right] \mathrm{d} t+g(t) \mathrm{d} \overleftarrow{\mathbf{w}}
\end{equation}
where $\overleftarrow{\mathbf{w}}$ is a backward Wiener process through the diffusion time.
In particular, the reverse process starts at $t = T$ and ends at $t = 0$. 

\section{GeCo and Fast-GeCo}
\label{sec:geco_fastgeco}
\subsection{GeCo}
\label{sec:geco}
The score function $\nabla_{\mathbf{x}_t} \log p_t\left(\mathbf{x}_t \mid \hat{\mathbf{s}}_1\right)$ is approximated by a neural network called score model $f_{\text{diff}}\left(\mathbf{x}_t, \hat{\mathbf{s}}_1, \mathbf{y}, t; \theta_{\text{diff}}\right)$, which is conditioned by the signal at the current state $\mathbf{x}_t$, the estimated signal from the separator $\hat{\mathbf{s}}_1$, the mixture noisy audio $\mathbf{y}$ and the time state $t$ , parameterized by $\theta_{\text{diff}}$.
Following~\cite{vincent2011connection},
the score model is fit to the score of the perturbation kernel,
\begin{equation}
\label{eq10}
\mathcal{L}_{\text{diff}}:=\Vert f_{\text{diff}}\left(\mathbf{x}_t, \hat{\mathbf{s}}_1, \mathbf{y}, t; \theta_{\text{diff}}\right)+\frac{\mathbf{z}_t}{\sigma(t)}\Vert_2^2
\end{equation}
where $\Vert \cdot \Vert_2^2$ is the $l2$-norm.
This is called denoising score matching (DSM)~\cite{vincent2011connection}.
For each training step,
we sample the estimated separated signal for one of the speakers $\mathbf{s}_1$, given the discriminative separator network, the corresponding signal,  the original mixture $\mathbf{y}$, and a timestep $t$ uniformly from $\left[t_\epsilon, T\right]$, where $t_\epsilon>0$ is a small value that assures numerical stability~\cite{richter2023speech}.
Then, we obtain $\mathbf{x}_t$ using~\eqref{eq5} and minimize the DSM loss with~\eqref{eq10}.

Once $\theta_{\text{diff}}$ is available, 
we can generate an estimate of clean speech signals from $\hat{\mathbf{s}}_1$ by solving the reverse SDE in~(\ref{eq9}).
To find numerical solutions for SDEs, the Euler-Maruyama (EuM) first-order method~\cite{song2021scorebased} can be employed to reduce the number of iterations.
The interval $[0, T]$ is partitioned into M equal subintervals of width $\Delta t=T / M$, which approximates the continuous formulation into the discrete reverse process $\left\{\mathbf{x}_T, \mathbf{x}_{T-\Delta t}, \ldots, \mathbf{x}_0\right\}$.
More specifically, the reverse process starts with $\mathbf{x}_{T} \sim \mathcal{N}\left(\hat{\mathbf{s}}_1, \sigma\left(T\right)^2 \mathbf{I}\right)$ and iteratively computes
% \begin{equation}
% \begin{aligned}
% \label{eq11}
% \hat{\mathbf{x}}_{t_{i-1}} &=\mathbf{x}_{t_{i}} + g\left(t_i\right) \sqrt{\Delta t} \mathbf{z}_{t_i} \\
% &+\left[-\mathbf{f}\left(\mathbf{x}_{t_{i}}, \hat{\mathbf{s}}_1\right)+g\left(t_i\right)^2 f_{\text{diff}}\left(\mathbf{x}_{t_{i}}, \hat{\mathbf{s}}_1, \mathbf{y}, t_i; \theta_{\text{diff}}\right)\right] \Delta t
% \end{aligned}
% \end{equation}
\begin{equation}
\begin{aligned}
\label{eq11}
&\hat{\mathbf{x}}_{(i-1)\Delta t} =\mathbf{x}_{i\Delta t} + g\left(i\Delta t\right) \sqrt{\Delta t} \mathbf{z}_{i\Delta t} \\
&+\left[-\mathbf{f}\left(\mathbf{x}_{i\Delta t}, \hat{\mathbf{s}}_1\right)+g\left(i\Delta t\right)^2 f_{\text{diff}}\left(\mathbf{x}_{i\Delta t}, \hat{\mathbf{s}}_1, \mathbf{y}, i\Delta t; \theta_{\text{diff}}\right)\right] \Delta t
\end{aligned}
\end{equation}
where %$i \in \{1, 2, \ldots, M+1\}$.
$i=M,M-1,\dots,2,1$.
The last iteration outputs $\hat{\mathbf{x}}_0 $ approximating the clean speech signal.
The reverse starting point $0<T^{\prime} \leq T$ can be
used for trading performance for computational speed.
Using a small $T^{\prime}$ may degrade the performance but can reduce the number of iterations.

\subsection{Fast-GeCo}
\label{sec:fastgeco}
There are several drawbacks of the discretization reverse process in Section~\ref{sec:geco}.
First, there are discretization errors both from the EuM method and the noise schedule. 
Second, as the reverse process starts with $T^{\prime}$ instead of $T$, there is a prior mismatch between the terminating forward distribution and the initial reverse distribution~\cite{lay2023reducing}.
Third, $M$ reverse steps make the inference much slower than a discriminative model. 
In addition,
GeCo is trained to estimate the noise added between two steps, which is different from the goal of extracting the clean signal from the noise mixture.
Therefore, it is hard to use loss objectives such as SI-SNR loss to optimize the signal quality.

To address the above issues,
we propose a fast generative correction method that finetunes the GeCo model into a single reverse step during inference.
The idea is to optimize a single reverse process directly from the start step $\mathbf{x}_{T^{\prime}}$ to the initial step $\mathbf{x}_{0}$ with an SI-SNR loss. 
To be more specific,
the one-step output is obtained by
\begin{align}
\label{eq12}
\hat{\mathbf{x}}_{0} &=\mathbf{x}_{T^{\prime}} + g\left(T^{\prime}\right) \sqrt{T^{\prime}} \mathbf{z}_{T^{\prime}} \nonumber\\
&+T^{\prime}\left[-\mathbf{f}\left(\mathbf{x}_{T^{\prime}}, \hat{\mathbf{s}}_1\right)+g\left(T^{\prime}\right)^2 f_{\text{diff}}\left(\mathbf{x}_{T^{\prime}}, \hat{\mathbf{s}}_1, \mathbf{y}, T^{\prime}; \theta_{\text{diff}}\right)\right]
\end{align}
which is \eqref{eq11}, with $M=1$ and $\Delta t=T'$.

 The SI-SNR loss function is then used to fine-tune the corrector parameters $\theta_{\text{diff}}$,
\begin{equation}
\label{eq13}
\mathcal{L}_{\text{cor}}:= -10 \log _{10} \frac{\Vert \frac{\langle\hat{\mathbf{x}}_{0}, \mathbf{s}_1\rangle \mathbf{s}_1}{\Vert\mathbf{s}_1\Vert^2} \Vert^2}{\Vert\hat{\mathbf{x}}_{0}-\frac{\langle\hat{\mathbf{x}}_{0}, \mathbf{s}_1\rangle \mathbf{s}_1}{\Vert\mathbf{s}_1\Vert^2}\Vert^2}
\end{equation}

\section{Experiments}

\subsection{Datasets}
To evaluate the performance of the proposed methods,
we trained models on the Libri2Mix noisy dataset~\cite{cosentino2020librimix},
which is simulated by the WHAM$!$ noise data~\cite{Wichern2019WHAM} and the Librispeech utterances~\cite{panayotov2015librispeech}.
It contains two training subsets with 212 hours (\textit{train-360}) and 58 hours (\textit{train-100}) of audio respectively.
The \textit{dev} set with 11 hours of audio was used for validation.
We then evaluated our models on the \textit{test} set, containing 11 hours of audio.

Furthermore, we evaluated these trained models on out-of-domain data. Utterances from the Wall Street Journal (WSJ) corpus were used as the speech source, while noise audio from WHAM$!$, MUSAN~\cite{snyder2015musan} and DEMAND~\cite{hadad2014multichannel} were used as the noise source for simulation. Each of these test sets has a duration of 5 hours.
To maintain consistency with the wsj0-2mix dataset \cite{hershey2016deep}, we ensured that the relative levels between the two speakers were preserved. Following the approach in \cite{Wichern2019WHAM}, noise was introduced by sampling a random Signal-to-Noise Ratio (SNR) value from a uniform distribution ranging from -6 to +3 dB. In addition, we generated a minimum-length version of the simulated data by removing any leading and trailing noise and truncating it to match the length of the shorter of the two speakers' utterances. All the generated mixtures were resampled to 8 kHz as done in the previous works \cite{luo2019conv, luo2020dual}.

\subsection{Baseline Methods}
Apart from GeCo and Fast-GeCo, 
we implemented the following baseline methods under the same experimental settings. \\
\noindent \textbf{SepFormer}~\cite{subakan2021attention}: is a discriminative model providing state-of-the-art results in Libri2Mix dataset, with one inference step. \\
\noindent \textbf{SepFormer-SE}: consists of a SepFormer and an enhancement model (an advanced model called SGMSE+~\cite{richter2023speech} was used here). SepFormer and SGMSE+ were trained independently. SGMSE+ was trained on the WHAM! dataset and needs 60 steps during the inference. \\
\noindent \textbf{SE-SepFormer}: is different from SepFormer-SE that SGMSE+ is used before SepFormer. \\
\noindent \textbf{DiffSep}~\cite{scheibler2023diffusion}: is a fully generative separation method based on diffusion with 30 inference steps. \\
\noindent \textbf{Refiner}~\cite{hirano2023diffusionbased}: is a combination of discriminative model and generative model.
As the number of inference steps is not clarified in the original paper, we chose an optimal number of 30 from our experiments. \\
\noindent \textbf{SpeechFlow}~\cite{liu2024generative} is a generative model followed by a discriminative model which requires a pre-training stage with 60k hours of audio data. 

\subsection{Metrics}
We evaluated the performance on reference-based perceptual metrics, including Perceptual Evaluation of Speech Quality improvement (PESQi)~\cite{rix2001perceptual},
Extended Short-Time Objective Intelligibility improvement (ESTOi)~\cite{jensen2016algorithm}, and
SI-SNR improvement (SI-SNRi).
We also introduced a reference-free metric (non-intrusive speech quality assessment (NISQA)~\cite{mittag21_interspeech},
which utilizes a deep neural network to estimate the mean opinion score (MOS) of a target signal without any reference signal.

\begin{table*}[t]
% \small
  \caption{Speech separation results on different test sets. All models are trained on Libri2Mix training set. $\star$ are the results of models reproduced by us. "SE" denotes a speech enhancement model. Values indicate mean and standard deviation.}
  \vspace{-3mm}
  \label{tab:result1}
  \centering
  %\customfontsize{5pt}{6.5pt}
  % \tiny
  \resizebox{\textwidth}{!}{
  \begin{tabular}{@{}l|lll|lll|lll|lll@{}}
    \hline
    \multirow{2}{*}{Method} & \multicolumn{3}{c|}{Libri2Mix} & \multicolumn{3}{c|}{WHAM!} & \multicolumn{3}{c|}{MUSAN} & \multicolumn{3}{c}{DEMAND}\\
    \cline{2-13}
    & PESQi & ESTOIi & SI-SNRi & PESQi & ESTOIi & SI-SNRi & PESQi & ESTOIi & SI-SNRi & PESQi & ESTOIi & SI-SNRi\\
    \hline
    SepFormer$\star$~\cite{subakan2021attention} & 0.59$\pm$0.25 & 0.23$\pm$0.06 & 10.58$\pm$2.63 & 0.40$\pm$0.26 & 0.20$\pm$0.08 & 9.91$\pm$3.72 & 0.50$\pm$0.36 & 0.20$\pm$0.10 & 9.06$\pm$4.83 & 0.67$\pm$0.34 & 0.21$\pm$0.09 & 9.89$\pm$4.89\\
    SepFormer-SE$\star$ & 0.72$\pm$0.30 & 0.24$\pm$0.07 & 9.80$\pm$2.44 & 0.45$\pm$0.26& 0.23$\pm$0.10 & 9.85$\pm$3.88 & 0.61$\pm$0.37& 0.21$\pm$0.08& 8.89$\pm$4.25& 0.82$\pm$0.30& 0.25$\pm$0.10& 9.98$\pm$4.70\\
    SE-SepFormer$\star$ & 0.46$\pm$0.22& 0.17$\pm$0.04& 7.60$\pm$2.05& 0.29$\pm$0.18& 0.14$\pm$0.05& 7.44$\pm$2.90& 0.35$\pm$0.19& 0.13$\pm$0.05& 7.20$\pm$2.77& 0.55$\pm$0.25& 0.18$\pm$0.06& 8.02$\pm$4.02\\
    DiffSep$\star$~\cite{scheibler2023diffusion} & 0.50$\pm$0.22 & 0.22$\pm$0.05 & 8.90$\pm$2.30 & 0.33$\pm$0.22 & 0.18$\pm$0.07 & 8.71$\pm$3.33 & 0.41$\pm$0.22 & 0.19$\pm$0.08 & 8.44$\pm$4.40& 0.62$\pm$0.30& 0.20$\pm$0.08& 9.47$\pm$4.53\\
    Refiner$\star$~\cite{hirano2023diffusionbased} & 0.64$\pm$0.25 & 0.23$\pm$0.05 & 10.10$\pm$2.70 & 0.42$\pm$0.26 & 0.21$\pm$0.09 & 9.74$\pm$3.66 & 0.51$\pm$0.35 & 0.20$\pm$0.10 & 8.83$\pm$4.80 & 0.60$\pm$0.30 & 0.20$\pm$0.09 & 9.55$\pm$4.54\\
    SpeechFlow~\cite{liu2024generative} & - & 0.33 & 10.46 & - & -& -& -& -&- & -& -& -\\
    \hline
    \textbf{GeCo (ours)} & 0.77$\pm$0.28 & 0.28$\pm$0.06 & 10.88$\pm$2.71 & 0.48$\pm$0.28 & 0.26$\pm$0.09 & 10.41$\pm$4.13 & 0.71$\pm$0.39 & 0.25$\pm$0.10 & 9.97$\pm$4.91& 0.88$\pm$0.37 & 0.25$\pm$0.10 & 9.85$\pm$4.77\\
    \textbf{GeCo-1step (ours)} & -0.26$\pm$0.11& -0.08$\pm$0.03& -3.11$\pm$1.30& -0.20$\pm$0.10& -0.07$\pm$0.03& -3.85$\pm$1.15& -0.28$\pm$0.09& -0.10$\pm$0.06& -4.20$\pm$1.44& -0.15$\pm$0.10& -0.10$\pm$0.04& -3.55$\pm$1.81\\
    \textbf{Fast-GeCo (ours)} & \textbf{0.86}$\pm$0.31 & \textbf{0.34}$\pm$0.07 & \textbf{12.98}$\pm$2.81 & \textbf{0.46}$\pm$0.28 & \textbf{0.32}$\pm$0.08 & \textbf{12.56}$\pm$3.59 & \textbf{0.75}$\pm$0.42 & \textbf{0.30}$\pm$0.11 & \textbf{12.29}$\pm$5.26 & \textbf{0.92}$\pm$0.33 & \textbf{0.29}$\pm$0.10 & \textbf{13.31}$\pm$4.91 \\
    \hline
  \end{tabular}}
  \vspace{-2mm}
\end{table*}

\begin{table}[t]
% \small
  \caption{NISQA results on different test sets.}
  \vspace{-3mm}
  \label{tab:result2}
  \footnotesize
  \centering
  \resizebox{\columnwidth}{!}{
  \begin{tabular}{@{}l|cccc@{}}
    \hline
    Method & Libri2Mix & WHAM! & MUSAN & DEMAND\\
    \hline
    Mixture & 1.15 & 0.96 & 1.64 & 1.99\\
    Clean & 3.47 & 3.24 & 3.24 & 3.24\\
    \hline
    SepFormer~\cite{subakan2021attention} & 2.05 & 1.79 &  2.23 & 2.35\\
    DiffSep~\cite{scheibler2023diffusion} & 2.44 & 2.13 & 2.30 & 2.98\\
    Refiner~\cite{hirano2023diffusionbased} & 3.46 & 3.60 & 3.33 & 3.75 \\
    \hline
    \textbf{GeCo (ours)} & 3.65 & 3.58 & \textbf{3.63} & 3.61 \\
    \textbf{Fast-GeCo (ours)} & \textbf{3.68} & \textbf{3.76} & 3.59 & \textbf{3.87} \\
    \hline
  \end{tabular}}
  \vspace{-3mm}
\end{table}

\subsection{Implementation Details}
SepFormer was used as our discriminative separator, following the same architecture as~\cite{subakan2021attention}. We set the first convolutional stride as 80,
which significantly speeded training up with minimal impact on performance.
We trained SepFormer on the Libri2Mix \textit{train-360} for 200 epochs set using
Adam optimizer with 16 batch-size, and initial learning rate of 0.00015 which was halved after 2 epochs without improvement.
%The SepFormer is trained for 200 epochs with a batch size of 16.

We further trained the GeCo and Fast-GeCo on the Libri2Mix \textit{train-100} set.
We employed the same noise conditional score network (NCSN++) architecture for the score matching function $f_{\text{diff}}$, as in SGMSE+~\cite{richter2023speech}. NCSN++ is based on multi-resolution U-Net structure with two input/output channels for the real and imaginary parts of the complex-valued Short-Time Fourier Transform (STFT).
Using the complex STFT allows us to obtain a more computationally efficient model compared to alternative diffusion models based on wave samples.
For GeCo, based on~\cite{lay2023reducing}, we set $T=0.999$, $t_\epsilon=0.03$, $k = 2.6$, $c = 0.51$ and the reverse starting point $T^\prime=0.5$. We take $M=30$ steps for the inference.
We used the Adam optimizer with a learning rate of 0.0001, 100 epochs, and a batch size of 32. 
An exponential moving average with a decay of 0.999 is used for the weights of the network~\cite{song2020improved}.
For Fast-GeCo, we trained for 50 epochs with the Adam optimizer with a learning rate of 0.0001 and batch size of 32.

All models were trained on two NVIDIA A100 GPUs, each equipped with 40 GB of GPU memory. We choose the best models and hyper-parameters based on the SI-SNR metrics from the validation set.

\section{Results and Analysis}

Speech separation results with reference-based metrics are presented in Table~\ref{tab:result1}.
On the Libri2Mix test set, the discriminative model, namely SepFormer, achieved a mean SI-SNRi of $10.58$.
Introducing an enhancement model at the end of the separation model (SepFormer-SE) improved PESQi and ESTOi metrics while using the enhancement model in front of the separation model led to a drop in performance. This suggests that the outputs of the enhancement model may introduce severe artifacts, which the separation model is not able to handle, or that the enhancement model is not suitable for overlapped audio. 

Compared to SepFormer, previous diffusion-based methods (DiffSep, Refiner, and SpeechFlow) did not significantly improve the metrics, particularly SI-SNRi. DiffSep, being a fully-generative model, struggles with the problem of speaker permutations during training, resulting in the worst performance. SpeechFlow requires a pre-training stage with 60k hours of English speech but failed to enhance the SI-SNRi metric. Refiner combines SepFormer with a denoising diffusion restoration model conditioned by SepFormer's output, yet it did not significantly improve the separation metrics, possibly because the original mixture speech is not considered as the conditional information of the diffusion model.

Similar to Refiner, our proposed method, GeCo, introduces a discriminative model for initial separation, addressing the issue of speaker permutations during training. Additionally, GeCo utilizes the original mixture speech as the conditional information of the diffusion reverse process, solving the Brownian Bridge SDE, and achieving better PESQi and SI-SNRi metrics than state-of-the-art methods.

However, these diffusion-based methods require multiple reverse steps for the inference stage, making them slower than SepFormer. Furthermore, as described in Section~\ref{sec:fastgeco}, discretization errors in the reverse process influence their performance, hindering significant metric improvements. By simply taking one reverse step in GeCo (GeCo-1step in Table~\ref{tab:result1}), the results were suboptimal. To address these issues, we fine-tuned GeCo into one reverse step with an SI-SNR loss, resulting in the proposed Fast-GeCo, which outperformed all other methods.

For out-of-domain data (WHAM!, MUSAN, and DEMAND), similar trends are observed, with GeCo and Fast-GeCo achieving superior results compared to previous models. Fast-GeCo demonstrates state-of-the-art performance in all experimental settings, indicating robust noise handling and generalization to out-of-domain data. Compared to SepFormer, Fast-GeCo improved PESQi by 15-50\%, ESTOIi by 38-60\% and SI-SNRi by 22-35\%.

\begin{figure}[t]
  \centering
  \includegraphics[width=\columnwidth]{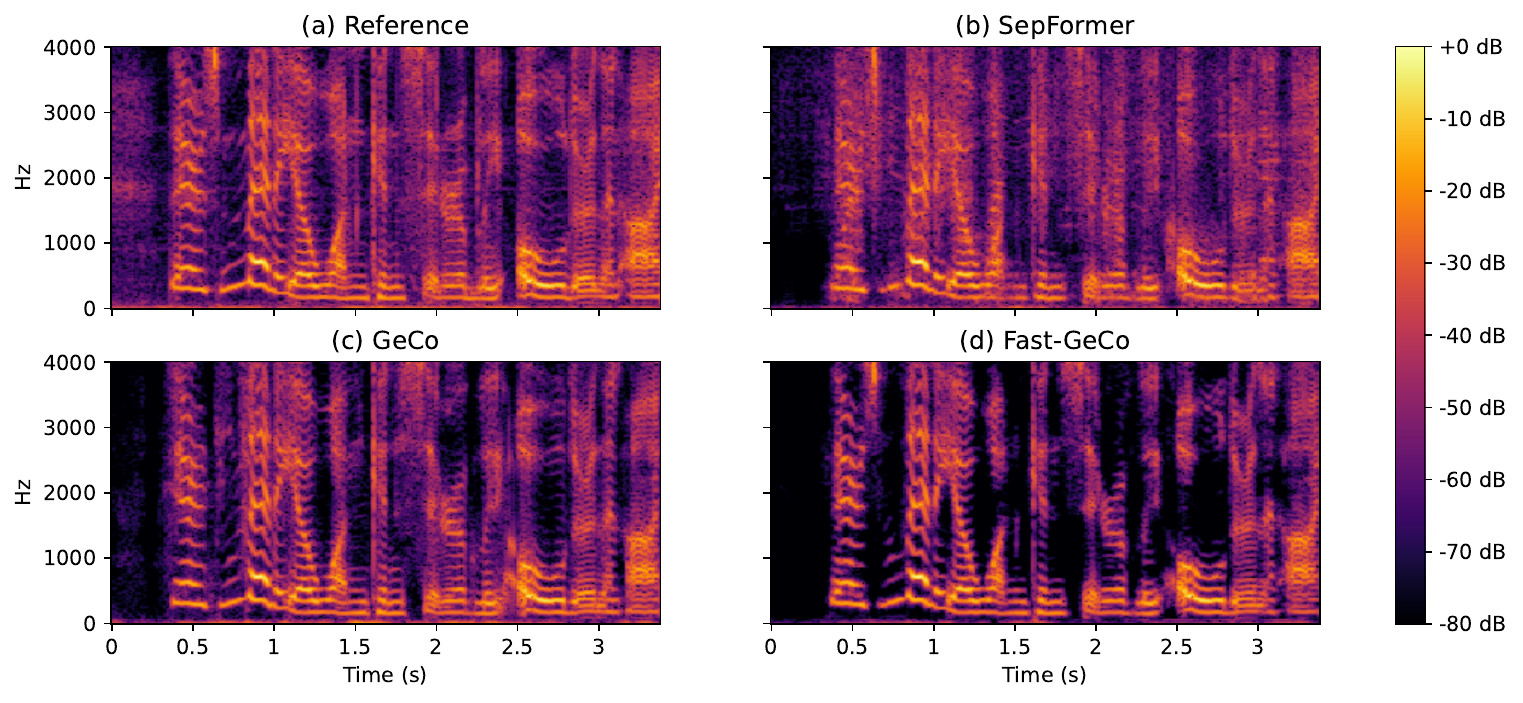}
  \vspace{-3mm}
  \caption{Mel spectorgram separated examples of Libri2Mix. (a) Reference. (b) Output of Sepformer. (c) Output of GeCo. (d) Output of Fast-GeCo.}
  \label{fig:mels}
  \vspace{-2mm}
\end{figure}

It is also crucial to consider reference-free metrics since the generative model's output may not align perfectly with the ground truth. NISQA results in Table~\ref{tab:result2} show significant improvements for all generative methods, with Fast-GeCo achieving the best results. Processed examples in Figure~\ref{fig:mels} illustrate enhanced harmonics structure clarity and sufficient suppression of non-speech ambient noise compared to the original SepFormer output. Generative methods can even outperform clean references in NISQA, likely due to light background noises in the reference audios, consistent with the cleaner output of generated samples shown in Figure~2.

\section{Conclusions}
In this paper, we proposed a generative corrector (GeCo) based on a Brownian Bridge SDE diffusion model to enhance the output of a discriminative speech separator achieving noise-robust speech separation. In order to speed up the multi-step diffusion reverse process, this corrector was further finetuned into a single-step reverse process (Fast-GeCo).
The proposed Fast-GeCo outperformed GeCo and other robust baselines achieving state-of-the-art results in both reference-based and reference-free metrics on the Libri2Mix noisy dataset. It also showed great generalization capabilities on the WSJ out-of-domain data.

We observed a limitation of this work, i.e., GeCo may encounter difficulties when processing outputs from SepFormer that exhibit significant confusion between the two speakers.
In future work, we will (i) design more effective and efficient backbones for the separator and the corrector, (ii) test the method in other tasks like target speech extraction, (iii) explore the method in the settings of more speakers.

{
\sevenhalfpt % Set the font size for the bibliography
\
\bibliographystyle{IEEEtran}
\bibliography{mybib}
}

\end{document}